\documentclass[aps,prx,twocolumn,showpacs,amsmath,amssymb,floatfix,superscriptaddress,notitlepage]{revtex4-2}
\usepackage[utf8]{inputenc}
\usepackage[left=2cm,right=2cm]{geometry}
\usepackage{physics}
\usepackage{graphicx}
\usepackage{xcolor}
\usepackage{tikz}
\usepackage{xcolor}

\usepackage{changes}

\usetikzlibrary{arrows.meta}
\usetikzlibrary{calc}
\usetikzlibrary{decorations.markings}
\usetikzlibrary{decorations}
\usetikzlibrary{positioning}

\begin{document}
\title{Variational quantum simulation using non-Gaussian continuous-variable systems}

\author{Paolo Stornati}
\thanks{These authors contributed equally \\ paolo.stornati@icfo.eu\\
federico.centrone@icfo.eu.}
\affiliation{ICFO-Institut de Ciencies Fotoniques, The Barcelona Institute of Science and Technology, Mediterranean Technology Park, Avinguda Carl
Friedrich Gauss, 3, 08860 Castelldefels, Barcelona, Spain}

\author{Antonio Acin}
\affiliation{ICFO-Institut de Ciencies Fotoniques, The Barcelona Institute of Science and Technology, Mediterranean Technology Park, Avinguda Carl
Friedrich Gauss, 3, 08860 Castelldefels, Barcelona, Spain}
\affiliation{ICREA - Institució Catalana de Recerca i Estudis Avan\c cats, Lluís Companys 23, 08010 Barcelona, Spain}

\author{Ulysse Chabaud}
\affiliation{DIENS, École Normale Supérieure, PSL  University, CNRS, INRIA, 45 rue d’Ulm, Paris, 75005, France}

\author{Alexandre Dauphin}
\affiliation{ICFO-Institut de Ciencies Fotoniques, The Barcelona Institute of Science and Technology, Mediterranean Technology Park, Avinguda Carl
Friedrich Gauss, 3, 08860 Castelldefels, Barcelona, Spain}

\author{Valentina Parigi}
\affiliation{Laboratoire Kastler Brossel, Sorbonne Universit\'{e}, CNRS, ENS-Universit\'{e} PSL, Coll\`{e}ge de France, 4 place Jussieu, F-75252 Paris, France}

\author{Federico Centrone}
\thanks{These authors contributed equally \\ paolo.stornati@icfo.eu\\
federico.centrone@icfo.eu.}
\affiliation{ICFO-Institut de Ciencies Fotoniques, The Barcelona Institute of Science and Technology, Mediterranean Technology Park, Avinguda Carl
Friedrich Gauss, 3, 08860 Castelldefels, Barcelona, Spain}

\begin{abstract}
    This work introduces a novel approach to quantum simulation by leveraging continuous-variable systems within a photonic hardware-inspired framework. The primary focus is on simulating static properties of the ground state of Hamiltonians associated with infinite-dimensional systems, such as those arising in quantum field theory. We present a continuous-variable variational quantum eigensolver  compatible with state-of-the-art photonic technology. The framework we introduce allows us to compare discrete and continuous variable systems without introducing a truncation of the Hilbert space, opening the possibility to investigate the scenarios where one of the two formalisms performs better. We apply it to the study of static properties of the Bose--Hubbard model and demonstrate its effectiveness and practicality, highlighting the potential of continuous-variable quantum simulations in addressing complex problems in quantum physics.
\end{abstract}

\maketitle

\section{Introduction} Quantum computing has always been fueled by the profound motivation to simulate ground state properties and dynamics of complex quantum systems. 
Indeed simulating inherently non-classical physical properties 
becomes practically unfeasible due to exponential scaling of the classical simulations~\cite{richard1981feynman}. 

Among the diverse platforms explored for quantum simulation purposes, photonics stands out as an exceptionally promising candidate~\cite{aspuru2012photonic,sparrow2018simulating,peruzzo2014variational, Somhorst_2023}.
Photons have long coherence times and can be easily manipulated using established photonic technologies. Notably, various
quantum advantage experiments have been recently conducted with photonic systems~\cite{aaronson2010computational,Zhong_2020,Zhong_2021,madsen2022quantum}.

The prevalent framework for quantum simulations is based on discrete-variable quantum information, which requires truncating the infinite-dimensional Hilbert space when dealing with continuous systems, such as quantum fields. These truncations introduce systematic errors in the simulation and require a theoretical framework to be studied. On the other hand, continuous-variable (CV) quantum information is a very natural framework to describe systems with infinite-dimensional Hilbert spaces. In CV systems, observables typically involve physical quantities like position, momentum, or field amplitudes, which can take  a continuous range of values.

The CV framework offers unique advantages in certain quantum information tasks, such as quantum computation, cryptography, and sensing: CV systems, such as optical modes, can exhibit deterministic, high-dimensional entanglement and can be manipulated with various tools, including passive optics, non-linear crystals, single-photon detectors and homodyne detectors~\cite{yokoyama2013ultra,roeland2022mode,roman2023spectrally,Killoran_2019}.

The field of CV quantum simulations has remained largely uncharted territory due to the intrinsic challenges associated with describing and controlling such complex systems~\cite{marshall2015quantum,deng2016continuous, Killoran_2019}. However, recent advancements in both theoretical and experimental techniques have reached a level of maturity that enables the exploration and resolution of these longstanding issues.

This work presents an innovative method for quantum simulations, formulating a variational quantum eigensolver (VQE)~\cite{peruzzo2014variational} within a photonic hardware-inspired CV framework. In this work, we focus on quantum states that can be written as a single product of ladder operations applied to a Gaussian state, whereas the exploration of more expressive ansatz is left for future work. The primary aim is to simulate the ground state of Hamiltonians. We use our method to explore the groundstate properties of a many-body system, namely the Bose--Hubbard model. We demonstrate the effectiveness and practicality of our proposed simulation methodology: ground states of Hamiltonians are successfully found under realistic experimental conditions, showcasing the potential of CV quantum simulations in addressing complex problems in quantum physics.

Notice that there has been some seminal works on quantum simulations \cite{marshall2015quantum,jha2023continuous} and variational algorithms \cite{Killoran_2019, bangar2023experimentally} within CV systems. However, these contributions depend on non-Gaussian gates currently beyond technological reach. In contrast, our approach seamlessly integrates into existing photonic platforms, making it readily applicable. Furthermore, we identify the computational complexity scaling in classical simulations and provide estimates of the experimental resources required for implementing our protocol physically. This allows researchers to assess whether their experimental setup meets minimal requirements to obtain a quantum advantage.

Additionally, we construct a framework that allows us to perform the classical simulation of the estimation of observables on non-Gaussian CV quantum states exactly, without requiring a truncation of the Hilbert space. This distinguishes our work from previous results in which the physical implementation was not possible and the classical simulation relied on the cutoff, making the system equivalent to a discrete-variable (DV) one.
 
The following section presents an overview of the CV techniques utilized in our framework. We begin by introducing CV quantum computing and the Gaussian/non-Gaussian dichotomy, where, similarly to  the DV Clifford/non-Clifford dichotomy, quantum processes involving only Gaussian states can be simulated efficiently. When non-Gaussian operations are added into the framework, the classical complexity of simulating the system increases drastically, paving the way to a possible quantum advantage.
The other sections present our results. We first outline our CV-VQE protocol and its implementation using state-of-the-art photonic technology. Then, we present our simulation results and perform a benchmark analysis using classical numerical techniques, like exact diagonalization of the truncated Hamiltonian.

\section{CV quantum computing} 
In quantum computing, it is crucial to assess if a task can be efficiently performed with classical resources. CV states fall into two categories: Gaussian and non-Gaussian. Gaussian computations, involving only Gaussian states described by classical probability distributions, resemble classical computing \cite{menicucci2006universal}, while non-Gaussian states are needed to unlock the full potential of CV states' infinite-dimensional Hilbert space.

\textbf{Gaussian states}---Gaussian states are those that can be generated from the vacuum by linear and quadratic Hamiltonians. They can be described by a Gaussian probability distribution in the phase space of their non-commuting quadrature operators $\hat x$ and $\hat p$. These operators are typically called position and momentum in reference to the harmonic oscillator terminology. 
The restriction to Gaussian states further implies that the quantum states are fully characterized by their first and second quadrature moments. In particular, in what follows we set the mean value of the quadratures to zero (which can always be ensured up to a displacement) and assume that these quantum states are fully described by their covariance matrix $V$. In Appendix~\ref{appendix:gauss} we describe the fundamental tools required to manipulate the Gaussian states and operations used in this work.

Gaussian states and operations have found applications in different areas of quantum physics and information~\cite{weedbrook2012gaussian,serafini2023quantum}, ranging from metrology~\cite{friis2015heisenberg,nichols2018multiparameter} to cryptography~\cite{navascues2006optimality,grosshans2003quantum}, from quantum networks~\cite{nokkala2018reconfigurable,centrone2021cost} to quantum batteries~\cite{friis2018precision,centrone2021charging}, from reservoir computing~\cite{nokkala2021gaussian} to quantum correlations~\cite{kogias2015quantification}. Nonetheless, quantum computations using only Gaussian states, operations, and measurements can be efficiently simulated on a classical computer, preventing any form of quantum computational advantage~\cite{bartlett2002efficient}. As a consequence, we need to extend the considered set of states and operations to the realm of non-Gaussianity~\cite{walschaers2021non}.

\textbf{Non-Gaussian statistics}---
We are interested in estimating the expectation value of some operator $\hat M$ on a non-Gaussian state $\hat\rho_{NG}$
\begin{equation}\label{eq:mean}
    \langle\hat M \rangle_{\hat\rho_{NG}}=\Tr[\hat M \hat\rho_{NG}].
\end{equation}
To compute this quantity for infinite-dimensional (or large-dimensional) systems, the standard approach is to impose a cutoff on the dimension of the associated density matrices. However, this procedure may lead to systematic errors, that can be quantified for specific models \cite{PhysRevD.100.034518}.  Avoiding this truncation, we consider the case where the observable can be written as a product of ladder operators, which is the case for many interesting physical Hamiltonians, whereas the state can be written as another product of ladder operators acting on an underlying Gaussian state $\hat\rho_G$:
\begin{equation}\label{eq:obs_state}
    \hat{M}=\prod_j \hat{a}_{j}^{\#}, \;\;\; \hat \rho_{NG}= \frac{1}{K} \prod_i\hat{a}_{i}^{\#}\hat \rho_G,
\end{equation}
where for each mode ${\#}\in\{\dag,\cdot\}$ indicates whether the ladder operator is of creation or annihilation type, respectively, and where
$K:=\Tr[\smash{\prod_i\hat{a}_{i}^{\#}\hat\rho_G}]$ is a normalization factor due to the non-unitarity of ladder operators. We show in App.~\ref{appendix:nongauss} how to analytically estimate Eq.\ (\ref{eq:mean}) for arbitrary non-Gaussian states.

Note that any operator can be approximated by a polynomial of the ladder operators \cite{Hartung_2019}. As a consequence of the linearity of the trace, we can express the expectation value of any operator on any quantum state as a linear combination of terms like Eq.\ (\ref{eq:obs_state}). In an optical setup, a linear combination of ladder operators can be implemented by  alternating layers of Gaussian unitaries and ladder operations, e.g. single photon additions and subtractions \cite{chabaud2021classical}. The proposed set-up, formed by a single layer of Gaussian operation and ladder operation is generalized in Appendix \ref{appendix:moreLayers} to the case of multiple layers.

 \textbf{CV Complexity}---While simulating Gaussian states is efficient, the computational complexity of non-Gaussian states was only recently explored with the introduction of the stellar rank, a discrete non-Gaussian measure~\cite{chabaud2020stellar,chabaud2023resources}. 
The concept of stellar rank introduces a hierarchy within normalized quantum states living in infinite-dimensional Hilbert spaces, commonly referred to as the stellar hierarchy. It is defined operationally as the number of  ladder operations required to engineer the state up to a Gaussian transformation. This hierarchy's existence has been established in both single-mode \cite{lachman2019faithful,chabaud2020stellar} and multimode scenarios~\cite{chabaud2022holomorphic}.
   
Within this hierarchy, pure states characterized by finite stellar ranks can be  transformed into pure states of finite stellar rank through Gaussian unitary operations. 
Conversely, states with infinite rank, exemplified by GKP \cite{gottesman2001encoding} or cat states, can be approximated with arbitrary precision using finite-rank states~\cite{chabaud2021certification}. 
     
It is commonly assumed that bounded-error quantum polynomial time (\textsf{BQP}) circuits can simulate CV quantum computations through discretization and Hilbert space truncation. However, formal results supporting this assumption remain limited \cite{tong2022provably}. Conversely, it has been demonstrated that adaptive rank-preserving quantum computations encompass \textsf{BQP}-complete computations~\cite{chabaud2022holomorphic}.
   
In the realm of classical algorithms for  simulating bosonic computations, the computational complexity scales with the stellar rank of both the input state and the measurement setup of the original computation~\cite{chabaud2023resources}. This implies that the measurement of observables on a bosonic state, spanning an exponentially large outcome space, can be approximately  simulated in a time complexity that scales exponentially with the total stellar rank of both the state and measurements. 

The theory of stellar rank naturally applies to our framework, where the complexity of the states we are able to simulate scales with the number of ladder operations we successfully implement. 

    \section{CV-VQE architecture}
\textbf{VQE}---VQE~\cite{Peruzzo_2014} is a promising algorithm for implementing quantum simulations on emerging noisy intermediate-scale quantum (NISQ) computers~\cite{Preskill_2018}. This hybrid quantum algorithm is used in quantum chemistry, quantum simulations, and optimization problems~\cite{Cerezo_2021_Nature_Reviews_Physics,dawid2022modern}. Ref.~\cite{Purification_based_mitigation_google} provides a roadmap for the hardware requirement to have a clear quantum advantage in variational quantum simulations. 
The objective is to find the ground state of a given physical system by minimizing the energy $E=\langle\psi|\hat H|\psi\rangle$, where $\hat H$ represents the Hamiltonian of the system.

\begin{figure}[t!]
    \centering
    \includegraphics[width=\columnwidth]{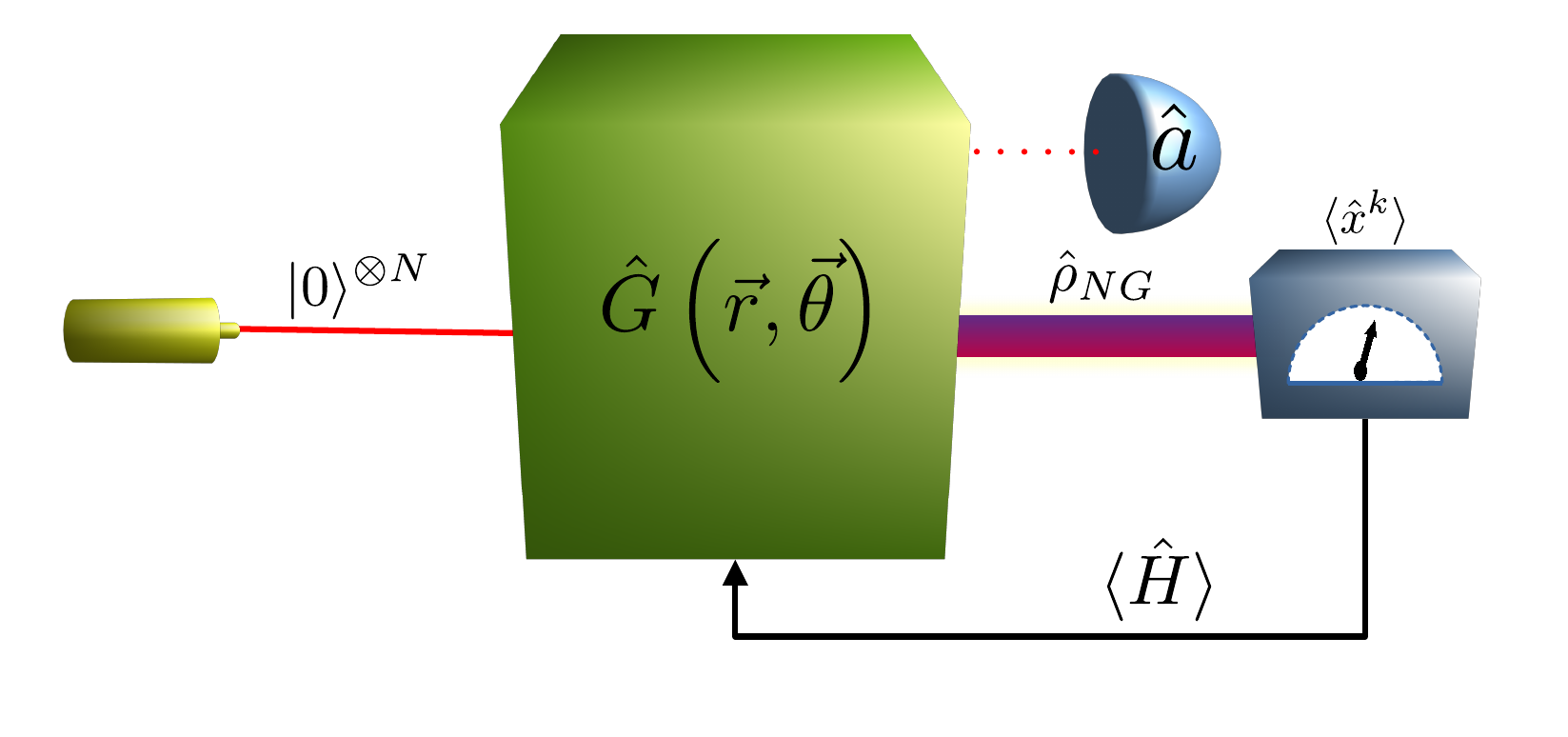}
    \caption{Photonic implementation of CV-VQE involves applying a general Gaussian unitary  $G(\vec r,\vec \theta)$, including multimode squeezing and passive optics, to the $N$-mode vacuum state generated by a laser. Conditioned on the success of the photon-subtraction,  measurements of field quadratures through homodyne detection enable energy estimation. Optimization techniques with a feedback loop are used to tune the Gaussian gate parameters and minimize the energy.}
    \label{fig:cvvqe_scheme}
\end{figure}

The VQE algorithm begins with the preparation of an initial state that can  easily be prepared on a quantum computer. Subsequently, a generic parametric unitary $U$, characterized by parameters $\vec\theta$, is applied to the initial state. The algorithm operates as a loop between a classical computer and a quantum computer (see Fig.\ \ref{fig:cvvqe_scheme}). The quantum hardware estimates the expectation value of the Hamiltonian, while the parameters are optimized to minimize the energy of the state.

Once the algorithm converges, an approximation of the system's ground state is obtained. To successfully reach the ground state, the parameterized quantum circuit, or ansatz, must be expressive enough to encompass the solution of the problem. However, if the ansatz is overly expressive, various issues may arise during the optimization process, such as barren plateaus  and high computational costs~\cite{McClean_2018}.

\textbf{Photonic implementation}---Our objective is to implement a VQE algorithm within a CV framework using a photonic quantum processor. To achieve a simulator that surpasses classical computational capabilities, the inclusion of non-Gaussian operations is essential \cite{bartlett2002efficient}. However, engineering such operations is challenging, so we aim to limit their usage based on current techniques. Ladder operations (e.g. single photon subtraction and addition) are readily available in most well-equipped  quantum photonics labs. These  are the simplest non-Gaussian operations  \cite{zavatta2008subtracting} and their effects and statistics are outlined in Appendix \ref{appendix:nongauss}.

With a high degree of experimental control \cite{nokkala2018reconfigurable,sansavini2019continuous}, the optimization process primarily focuses on the parameters of Gaussian operations, such as squeezing and passive optics. The scheme of our photonic CV-VQE algorithm is illustrated in Fig.\ \ref{fig:cvvqe_scheme}. Initially, a set of $N$-mode vacuum states are subjected to squeezing by non-linear crystals with squeezing factors $\vec r$. Subsequently, the modes undergo interference in a linear optical circuit with parameters $\vec\theta$, resulting in entanglement between the modes. Finally, a photon-subtraction operation is applied to a specific mode, and all other modes are measured using homodyne detection \cite{Kovalenko21,Ansquer21,Asa-multiplex-21,Koadou23}. The moments of the quadrature distributions obtained from these measurements encode the energy of the simulated Hamiltonian. By estimating these quantities, we can implement a classical optimization protocol to optimize the $N^2+N$ parameters of the general Gaussian unitary $G(\vec r, \vec \theta)$, minimizing the energy of the system, converging to an approximation of the ground state of the system \cite{yao2023design}.

Although our proposed VQE draws inspiration from particular hardware implementations, the versatility of our framework can be smoothly adapted to diverse experimental platforms that employ different natural operations. For instance, leveraging the mathematical tools presented in the appendix, we can describe a scheme wherein the initial state comprises a product of single-photon states going through a passive interferometer before being measured by homodyne detectors \cite{chabaud2021classical}. This showcases the adaptability and broad applicability of our approach beyond specific hardware configurations.

In the work by Bangar et al. \cite{bangar2023experimentally}, a photonic scheme akin to ours is developed, employing multiple layers of Gaussian and non-Gaussian gates to construct a CV quantum neural network. Notably, they emulate the approach outlined in \cite{Killoran_2019}, specifically detailing the implementation of the cubic phase gate through a sequence of three photon subtractions \cite{marshall2015repeat}. While this scheme is theoretically feasible, it remains impractical with current technology. Each layer of their neural network needs three probabilistic photon subtractions, whereas our approach requires only one to conduct non-trivial simulations efficiently.

Moreover, their classical simulations use number-resolving detectors instead of the more conventional threshold detectors. They also impose a Hilbert space cutoff in their simulations, whereas our newly introduced framework avoid such a truncation.

Additionally, the quantum scheme in \cite{bangar2023experimentally} employs a universal set of quantum gates, rendering it equivalent to a universal quantum computer. In contrast, our approach relies on Near-Term Intermediate-Scale Quantum (NISQ) operations, yet retains the potential to approximate the scheme outlined in \cite{Killoran_2019}. Finally, as detailed in the subsequent section, we provide realistic estimates for the experimental resources and time required to implement our protocol, enhancing the practicality and feasibility of our proposed approach.

 \textbf{Experimental resources}---In our approach, we use CV states with a high degree of purity, that are not far from reach particularly in optical system~\cite{Asavanant19,Larsen2019,Ra19}. To effectively implement this setup, the primary resources in demand are squeezing and single-photon operations. \begin{figure}[t!]
    \centering
    \includegraphics[width=0.9\columnwidth]{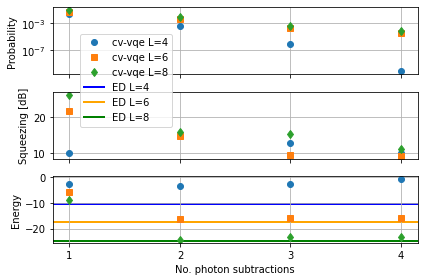}
    \caption{Simulation of the Bose--Hubbard hamiltonian with parameters $U=t=\mu=1$ and number of sites $L=4,6,8$. The ansatz states have a purity of $95\%$. We show the scaling of the probability of photon subtractions, the squeezing cost, and the energy of the ground state with the number of photon subtractions.}
    \label{fig:expResource}
\end{figure}

The degree of achievable squeezing is constrained by the non-linearity within SPDC crystals. This constraint can be quantified and addressed using the squeezing cost introduced in \cite{centrone2021cost}.
The amount of squeezing needed for the implementation depicted in Fig. \ref{fig:expResource} may pose challenges for schemes reliant on more than 10 frequency modes \cite{Cai17,Roman23}. Nonetheless, it remains feasible for temporal modes, up to approximately 3-5 dB \cite{madsen2022quantum}. While this could result in longer computational times, there is potential for improvement through a hybrid approach that combines frequency and temporal modes \cite{Koadou23,Roman23}.

It is worth noting that the probability of successfully implementing photon subtractions, as shown in Fig. \ref{fig:expResource}, decreases exponentially with the number of subtractions. However, considering the CV states generation rate, that in setup based on mode-locked lasers can be of the order of $100$ MHz,
 this task can be accomplished within a reasonable timeframe, on the order of  hundreds of seconds.

Furthermore, it is important to emphasize that the protocol does not necessitate full-state tomography. Instead, it only requires the estimation of the mean values of observables through homodyne detection. This is a standard practice for quadratic observables. However, it becomes particularly crucial when dealing with observables that are quartic in the quadratures~\cite{PhysRevA.96.053835}. 

\section{simulation of strongly correlated quantum many-body system}

First principle simulations of strongly correlated quantum many-body systems are crucial for the understanding of many physical phenomena \cite{lewenstein2007ultracold}. Firstly, it enhances our understanding of fundamental physics by studying the behavior of matter at the microscopic level. Secondly, it helps to design and discover materials with desired properties, such as high-temperature superconductors or topological phases of matter~\cite{RevModPhys.86.153}.

\textbf{Bose--Hubbard model in 1+1 d}---The model we consider within our variational framework is the Bose--Hubbard (BH) model in 1+1 dimension. Classical simulations of the model (e.g. with Exact Diagonalization ED or Density Matrix Renormalization Group DMRG \cite{PhysRevLett.69.2863}) are always performed considering a cutoff in the maximal occupation number of bosons per site. 
The Hamiltonian is given by:

\begin{equation}
    \begin{aligned}
        \hat H&=-t\sum_i(\hat b_i \hat b^\dag_{i+1}+\hat b^\dag_i \hat b_{i+1})\\
        &+\frac U2\sum_i\hat n_i\hat n_i-\left(\mu+\frac U2\right)\sum_i\hat n_i,
    \end{aligned}
\end{equation}
where $\mu$ is the chemical potential, $t$ is the hopping strength, $U$ is the on-site interaction magnitude, and $\hat{n}_i=\hat b^{\dagger}_{i}\hat b_{i}$.

\begin{figure}[t!]
    \centering
    \includegraphics[width=1.1\columnwidth]{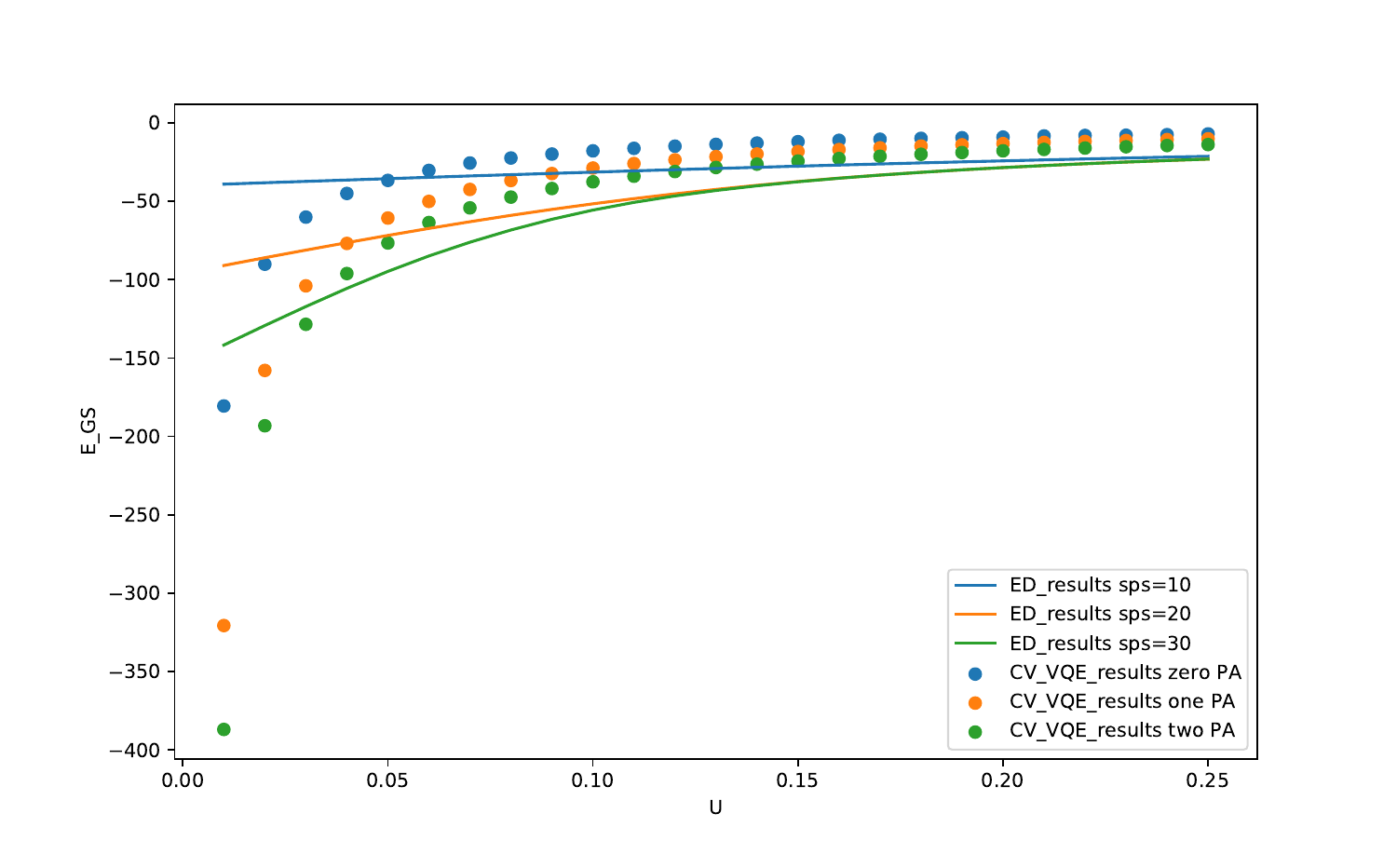}
    \caption{Scaling of the energy as a function of the on-site interaction magnitude $U$. The energy is minimized with the L-BFGS-b algorithm \cite{cite-key}. The exact diagonalization is computed with quspin and we put a cut-off of particles per site at 10, 20 and 30. We see that  the variational energy is lower than the one computed with a truncation of the Hilbert space for small values of $U$. PA stands for photon addition.}
    \label{fig:scaling_U_Addition}
\end{figure}

\textbf{Numerical results on Bose--Hubbard model}---We simulate variationally the BH model with CV-VQE. Gaussian ansätze have been used to simulate classically the BH model,\cite{Gaussian_variational_BH_cirac} where the authors developed an efficient Gaussian tensor network algorithm for high performance classical simulation. As described above, we have as variational parameters the squeezing parameters and the parameters of the linear optical network. For $N$ modes, we have in total $N^2+N$ variables.
In Fig.\ \ref{fig:scaling_U_Addition} the energy of the ground state is computed when we vary the on-site interaction strength $U$. For very small $U$, we see a consistent deviation from the values of the exact diagonalization computed with a cut-off of up to 30 bosons per site. The values of the CV-VQE are much lower than the ED solution performed with Quspin \cite{quspin_scipost} at a fixed cut-off. The numerical simulation are performed in the gran canonical ensemble without fixing the particle number. The values of $\mu$ and $t$ are fixed to 1. We observe that the energy of the ground state is unbounded from below for $U=0$, as we permit an infinite accumulation of particles per site. 

\section{Conclusion}\label{conclusion}
In summary, our work introduces a novel approach to quantum simulations utilizing continuous variables within a photonic hardware-inspired VQE framework. We demonstrate the framework's versatility through the exploration of the study of a paradigmatic model in  many-body physics. Crucially, our simulations were conducted under realistic experimental conditions. To test the robustness of our model, we considered constraints like impurity, squeezing cost, and the probability of single-photon operations.  Moreover, recent understanding of barren plateaus \cite{Ragone:2023qbn} do not directly apply to this framework, as we operate within an infinite dimensional Hilbert space. The extension of this study to the CV case will be addressed in a future study. 

We also showcase the framework's practicality by successfully identifying the ground states of Hamiltonians, even in the presence of realistic experimental constraints. Such framework, unlike previous works relying on Hilbert space truncation in the classical simulation, allows for the first time to classically simulate the potential advantage of CV systems in certain practical scenarios. Bridging the gap between theory and experiment, this gives valuable insights into quantum simulations. This novel approach not only opens new avenues for exploring complex quantum systems but also has the potential to advance CV quantum computing and information processing, suggesting which photonic technologies should be improved to achieve an advantage for the simulation of a quantum system.

\section{Acknowledgements}
We thank Todor Krasimirov, Egle Pagliaro and Marcin Plodzien for fruitful discussions.
P.S.\ acknowledges support from: ERC AdG NOQIA; MICIN/AEI (PGC2018-0910.13039/501100011033,  CEX2019-000910-S/10.13039/501100011033, Plan National FIDEUA PID2019-106901GB-I00, FPI; MICIIN with funding from European Union NextGenerationEU (PRTR-C17.I1): QUANTERA MAQS PCI2019-111828-2); MCIN/AEI/10.13039/501100011033 and by the “European Union NextGeneration EU/PRTR"  QUANTERA DYNAMITE PCI2022-132919 (QuantERA II Programme co-funded by European Union’s Horizon 2020 programme under Grant Agreement No 101017733), Proyectos de I+D+I “Retos Colaboración” QUSPIN RTC2019-007196-7); Fundació Cellex; Fundació Mir-Puig; Generalitat de Catalunya (European Social Fund FEDER and CERCA program, AGAUR Grant No. 2021 SGR 01452, QuantumCAT \ U16-011424, co-funded by ERDF Operational Program of Catalonia 2014-2020); Barcelona Supercomputing Center MareNostrum (FI-2023-1-0013); EU Quantum Flagship (PASQuanS2.1, 101113690); EU Horizon 2020 FET-OPEN OPTOlogic (Grant No 899794); EU Horizon Europe Program (Grant Agreement 101080086 — NeQST), National Science Centre, Poland (Symfonia Grant No. 2016/20/W/ST4/00314); ICFO Internal “QuantumGaudi” project; European Union’s Horizon 2020 research and innovation program under the Marie-Skłodowska-Curie grant agreement No 101029393 (STREDCH) and No 847648  (“La Caixa” Junior Leaders fellowships ID100010434: LCF/BQ/PI19/11690013, LCF/BQ/PI20/11760031,  LCF/BQ/PR20/11770012, LCF/BQ/PR21/11840013). 
U.C. and V.P. \ acknowledges funding from the European Union’s Horizon Europe Framework Programme (EIC Pathfinder Challenge project Veriqub) under Grant Agreement No.\ 101114899.  V.P. acknowledges support from ERC grant COQCOoN (Grant No.  820079)
F.C.\ acknowledges the Government of Spain (Severo Ochoa CEX2019-000910-S and European Union NextGenerationEU PRTR-C17.I1) and European Union (PASQuanS2.1, 101113690), Fundació Cellex, Fundació Mir-Puig, Generalitat de Catalunya (CERCA).
Views and opinions expressed are, however, those of the author(s) only and do not necessarily reflect those of the European Union, European Commission, European Climate, Infrastructure and Environment Executive Agency (CINEA), nor any other granting authority.  Neither the European Union nor any granting authority can be held responsible for them.

\bibliographystyle{ieeetr}
\bibliography{biblio}
\newpage
\appendix
\section{Gaussian states and operations}\label{appendix:gauss}

The canonical commutation relations between bosonic field operators, e.g.\ the harmonic oscillator ladder operators, read
\begin{equation}
    [\hat a_j, \hat a^{\dagger}_k]=\delta_{jk},
\end{equation}
where $\delta_{jk}$ is Kronecker's symbol. This relation implies that the quantum systems on which these operators act are described using infinite-dimensional Hilbert spaces \cite{serafini2023quantum}, often leading to cumbersome calculations. A typical restriction to make CV systems more tractable is to consider only Gaussian states and operations \cite{weedbrook2012gaussian}. The Gaussian regime applies when the initial state is the vacuum and the evolution is governed by Hamiltonians that are at most quadratic in the field operators. This restriction allows us to describe the states as a Gaussian probability distribution in the phase space of their quadratures. The quadratures operators $\hat x$ and $\hat p$, typically called position and momentum in reference to the harmonic oscillator terminology, correspond to the real and imaginary parts of the field operators
\begin{equation}\label{eq:ladders}
    \begin{cases}
    \hat x_j= \hat a_j +\hat{a}_j^\dagger\\
    \hat p_j= i(\hat{a}_j^\dagger-\hat a_j )
    \end{cases}\quad[\hat x_j, \hat p_k]=2i \delta_{jk}.
\end{equation}
The Gaussian restriction further implies that the quantum states are fully characterized by their first and second quadrature moments. In particular, in what follows we neglect the mean value of the quadratures and assume that quantum states are fully described by their covariance matrix $V$, defined as
\begin{equation}
    V=\Re\langle\Vec{\hat\xi}\;\Vec{\hat\xi}^T\rangle-\langle\Vec{\hat\xi}\rangle\langle\Vec{\hat\xi}^T\rangle=\Re\langle\Vec{\hat\xi}\;\Vec{\hat\xi}^T\rangle,
\end{equation}
where in the second equality we used the fact that the first moments are null and defined the vector of quadratures for a system of $N$ modes as $\Vec{\hat\xi}^T:=(\hat x_1, \dots, \hat x_N, \hat p_1,\dots, \hat p_N)$. Using the singular value decomposition (also called ``Bloch--Messiah'') of the symplectic transformations acting on the statistical moments of the quadratures, we can rewrite the covariance matrix as
\begin{equation}
    V=OZO^T,
\end{equation}
where $O\in Sp_{2N,\mathbb{R}}\cap SO(2N)\cong U(N)$ represents a passive transformation that commutes with the number operator $\hat n=\hat a^\dag\hat a$ and $Z=\mathrm{diag}(r_1,...,r_N,1/r_1,...,1/r_N)$, represents a squeezing transformation that  reduces the variance of the position of mode $j$ by a factor $0<r_j\leq 1$, while the complementary momentum operator is antisqueezed by the same factor.

Gaussian states and operations have found applications in different areas of quantum physics and information \cite{weedbrook2012gaussian,serafini2023quantum}, ranging from metrology \cite{friis2015heisenberg,nichols2018multiparameter} to cryptography \cite{navascues2006optimality,grosshans2003quantum}, from quantum networks \cite{nokkala2018reconfigurable,centrone2021cost} to quantum batteries \cite{friis2018precision,centrone2021charging}, from reservoir computing \cite{nokkala2021gaussian} to quantum correlations \cite{kogias2015quantification}. Nonetheless, computation with a Gaussian restriction can be efficiently simulated on a classical computer, preventing any form of computational quantum advantage \cite{bartlett2002efficient}. As a consequence, we need to extend our sets of operations to the realm of non-Gaussianity \cite{walschaers2021non}.

\section{Non-Gaussian statistics}\label{appendix:nongauss}

As explained in the main text, we are interested in estimating the expectation value of some operator $\hat M$ on a non-Gaussian state $\hat\rho_{NG}$
\begin{equation}
    \langle\hat M \rangle_{\hat\rho_{NG}}=\Tr[\hat M \hat\rho_{NG}].
\end{equation}
To compute this quantity for the infinite-dimensional systems we are interested in, the standard approach would be to impose a cutoff to the dimension of the associated matrices. To avoid this approximation, we can assume instead that the observable can be written as a product of ladder operators, whereas the state can be written as another product of ladder operators acting on an underlying Gaussian state $\hat \rho_G$:
\begin{equation}
    \hat{M}=\Pi_j \hat{a}_{j}^{\#}, \;\;\; \hat \rho_{NG}= \frac{1}{K} \Pi_i\hat{a}_{i}^{\#}\hat \rho_G,
\end{equation}
where for each mode ${\#}\in\{\dag,\cdot\}$ indicates whether the field operator is of creation or annihilation type, respectively, and where
$K:=\Tr[\smash{\Pi_i\hat{a}_{i}^{\#}\rho_G}]$ is a normalization factor that comes from the fact that ladder operators are non-unitary.

Let $\hat\rho_G$ be the density matrix of a $N$-mode Gaussian quantum state.
We know from Wick's theorem (see also \cite{walschaers2017statistical,walschaers2020emergent}) that we can compute the expectation value of products of ladder operators on a Gaussian state as
\begin{align}\label{eq:stats}
    \nonumber\Tr [\hat{a}^{\dagger}_{S_n}...\hat{a}^{\dagger}_{S_1}\hat{a}^{\dagger}_{C_m}...\hat{a}^{\dagger}_{C_1}\hat{a}_{C_1}...\hat{a}_{C_m}\hat{a}_{S_1}...\hat{a}_{S_n}\hat\rho_G]\\
    =\sum_{\mathcal{P}}\prod_{\{(p_1,\#),(p_2,\#)\}\in  \mathcal{P}}\Tr[\hat{a}_{p_1}^{\#}\hat{a}_{p_2}^{\#}\hat\rho_G],
\end{align}
where $\mathcal{P}$ is the set of all possible perfect matchings of the $2n+2m$ indices $(p_k,\#)\in\{C_1,\dots,C_m,S_1,\dots,S_n\}\times\{\dag,\cdot\}$ of the ladder operators. These operators are divided into those implementing the photon-subtraction on mode $j$ of the state ($\hat{a}_{S_j}$) and those indicating the observable on mode $j$ of which we wish to infer the statistics ($\hat{a}_{C_j}$). Notice that mathematically these are the same object and thus implementing more ladder operations is equivalent  to looking at the statistics of the higher-order moments of the ladder operators, albeit up to a normalization factor (which can be computed using the same formula).

Note as well that for an initial Gaussian state with no displacement, all the odd order correlations vanish \cite{walschaers2017statistical}, thus we always work with an even number of ladder operators. The pair $\{(p_1,\#),(p_2,\#)\}$ indicates the modes of the two operators in the product, e.g.\ for $\{(p_1,\#),(p_2,\#)\}=\{(C_2,\cdot),(S_3,\dag)\}$ we have $\Tr[\smash{\hat{a}_{p_1}^{\#}\hat{a}_{p_2}^{\#}\hat\rho_G}]=\Tr[\smash{\hat{a}_{C_2}\hat{a}_{S_3}^{\dagger}\hat\rho_G}]$. All the elements of the product of Eq.\ (\ref{eq:stats}) can be computed by using the following identities:
\begin{align} 
    I_1&=\Tr[\hat{a}_{j}^{\dagger}\hat{a}_{k}^{\dagger}\hat\rho_G]\label{eq:identity1}\\
   \nonumber&=\frac{1}{4}\left[V_{jk}-V_{j+N,k+N}-i(V_{j,k+N}+V_{j+N,k})\right],\\
   I_2&= \Tr[\hat{a}_{j}\hat{a}_{k}\hat\rho_G]=I_1^*\label{eq:identity2}\\
    I_3&=\Tr[\hat{a}_{j}^{\dagger}\hat{a}_{k}\hat\rho_G]\label{eq:identity3}\\
   \nonumber&=\frac{1}{4}\left[V_{jk}+V_{j+N,k+N}+i(V_{j,k+N}-V_{j+N,k})-2\delta_{jk}\right],\\
   I_4&=\Tr[\hat{a}_{j}\hat{a}^{\dagger}_{k}\hat\rho_G]\label{eq:identity4}=\delta_{jk}+I_3
\end{align}
where $V_{jk}$ is the $(j,k)$ matrix element of the $2N\times2N$ covariance matrix of the quadratures  of the state $\hat\rho_G$.

Wrapping up in a single formula we obtain
\begin{equation}\label{eq:expVal}
    \begin{aligned}
      \langle M \rangle_{\rho_{NG}}&=\frac1K\Tr[\Pi_j\hat a_j^{\#}\Pi_i\hat a_i^{\#}\hat\rho_G]\\
      &=\frac1K\sum_{\mathcal{P}}\prod_{\{(p_1,\#),(p_2,\#)\}\in  \mathcal{P}}\Tr[\hat a_{p_1}^{\#}\hat a_{p_2}^{\#}\hat\rho_G],
    \end{aligned}
\end{equation}
which is just a sum of the product of the identities in Eqs.\ (\ref{eq:identity1}-\ref{eq:identity4}) that arise from the perfect matching. The latter is the most computationally demanding operation, as it scales like $N!!$. While there exists optimized algorithms that reduce this cost to exponential in $N$ \cite{bjorklund2012counting}, this is at the origin of the quantum advantage in Gaussian boson sampling \cite{hamilton2017gaussian,chabaud2017continuous}. 

Note that any operator can be approximated by a polynomial of the ladder operators \cite{Hartung_2019}. As a consequence of the linearity of the trace, we can express the expectation value of any operator on any quantum state as a linear combination of terms like Eq.\ (\ref{eq:expVal}). In an optical implementation, this would correspond to alternating layers of Gaussian unitaries and ladder operations \cite{chabaud2021classical}. In this work, we focus on quantum states that can be written as a single product of ladder operations applied to a Gaussian state, whereas the exploration of more expressive ansatz is left for future work.
This generalized framework is discussed in Appendix \ref{appendix:moreLayers}.

\section{Multiple layers}\label{appendix:moreLayers}

 \begin{figure}[h]
    \centering
    \includegraphics[width=\columnwidth]{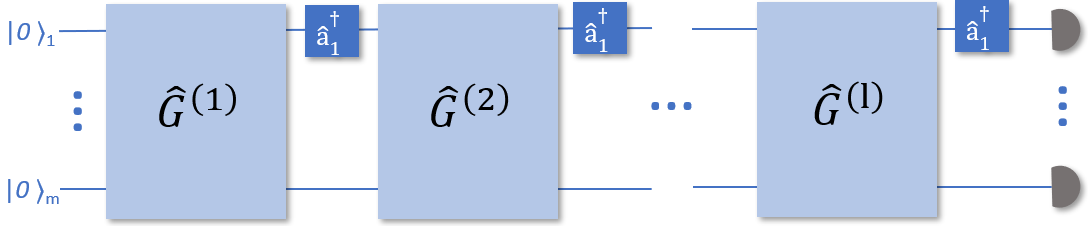}
    \caption{IPAG circuit with $l$ layers and $m$ modes.}
    \label{fig:IPAG}
\end{figure}

The single-layer VQE is not expressive enough to encompass the diversity of CV states. As a consequence, in some cases adding many single photon operations might lead further from the target ground state (as in Fig.\ \ref{fig:expResource}). A simple generalization of this framework is to add more layers and implement an interleaved
photon-added Gaussian  (IPAG) circuit, shown in Fig.\ \ref{fig:IPAG}. These types of CV quantum circuits were studied in \cite{chabaud2021classical} and are known to be classically simulable when the number of ladder operations grows logarithmically with the number of modes \cite{bourassa2021fast}. 

After $l$ layers of the IPAG circuit, the density matrix of the final state will be of the form:

\begin{equation}
\label{multilayer}
    \hat{a}_1^{\dagger} \hat{G}^{(l)}
    \hat{a}_1^{\dagger}  ... \hat{G}^{(2)}\hat{a}_1^{\dagger} \hat{G}^{(1)} \  
    \ket{0}\bra{0}^{\otimes N}
    \hat{G}^{\dagger(1)} \hat{a}_1 \hat{G}^{\dagger(2)}...
     \hat{a}_1
    \hat{G}^{\dagger(l)} \hat{a}_1.
\end{equation}

 To compute the expectation value of some operator on this state we would like to use Eq.\ (\ref{eq:expVal}) as in the single layer case. To do so, we need all the ladder operators to act directly on a Gaussian state. For that to happen we must commute the Gaussian unitaries with the ladder operators for them to be applied to the initial multimode vacuum state. The effect of this commutation follows from the relation 

 \begin{equation}
\label{unit_act_gaus_lad}
    \hat{G}\hat{a}_k^\dagger = \left(\sum_{j=1}^N s_{k,j}\hat{a}_j^\dagger + s_{k,N+j}\hat{a}_j\right)\hat{G} = \mathbb{S}_k\hat{G},
\end{equation}
where $\mathbb{S}_k$ is a superposition of ladder operators for each mode $j\in{1,..., N}$ with coefficients $s_{j,k}\in \mathbb{R}$ corresponding to elements of the symplectic matrix associated to the unitary matrix $\hat{G}$. Plugging this into Eq.\ (\ref{multilayer}) yields the  state in the desired form

\begin{equation}
\label{reduced-commutated}
    \hat{a}_1^\dagger
    \mathbb{S}^{(l-1)}...\ \mathbb{S}^{(1)}
    \hat{G} \ 
    \ket{0}\bra{0}^{\otimes N}
    \hat{G}^\dagger
    \mathbb{S}^{\dagger(1)}...\ \mathbb{S}^{\dagger(l-1)}
    \hat{a}_1,
\end{equation}
where the product of all the Gaussian unitaries was incorporated into a global unitary $\hat G$ acting on the vacuum. We can now employ Eq.\ (\ref{eq:expVal}), that will produce linear combinations of expressions of the form of Eq.\ (\ref{eq:stats}). The coefficients of such a combination will be optimized in the training process together with the parameters of the global Gaussian operation $\hat G$.

By including enough layers, this setup can approximate a very large family of quantum states with arbitrary precision. Additionally, the superposition of creation and annihilation operators allows us to interpolate between photon subtractions and additions, letting the optimization find the best choice and increasing the expressivity of our VQE. We will employ this framework to construct a CV neural network in future work.

\end{document}